\documentclass[aps,pra,twocolumn,superscriptaddress,bibliography]{revtex4-1}

\usepackage{graphicx}
\usepackage{graphics}
\usepackage{dsfont}
\usepackage{bm}
\usepackage{amsmath}
\usepackage{amsfonts}
\usepackage{mathrsfs}
\usepackage{braket}
\usepackage[capitalize]{cleveref}
\usepackage{units}
\usepackage{color}

\newcommand{\cop}{\hat{c}^{}}

\newcommand{\Hop}{\hat{H}}
\newcommand{\copd}{\hat{c}^{\dagger}}

\Crefname{fig.}{Fig.}{Fig.}
\Crefname{section}{Sec.}{Sec.}

\begin{document}

\title{Coulomb effects on the photo-excited quantum dynamics of electrons in a plasmonic nanosphere}

\author{Alexandra Crai}
\email[]{alexandra.crai14@imperial.ac.uk}
\author{Andreas Pusch}
\affiliation{Department of Physics, Imperial College London, London SW7 2AZ, UK }
\author{Doris E. Reiter}
\affiliation{Department of Physics, Imperial College London, London SW7 2AZ, UK }
\affiliation{Institut f\"ur Festk\"orpertheorie, Universit\"at M\"unster, Wilhelm-Klemm-Strasse 10, 48149 M\"unster, Germany}
\author{Lara Rom\'an Castellanos}
\affiliation{Department of Physics, Imperial College London, London SW7 2AZ, UK }
\author{Tilmann Kuhn}
\affiliation{Institut f\"ur Festk\"orpertheorie, Universit\"at M\"unster, Wilhelm-Klemm-Strasse 10, 48149 M\"unster, Germany}
\author{Ortwin Hess}
\email[]{o.hess@imperial.ac.uk}
\affiliation{Department of Physics, Imperial College London, London SW7 2AZ, UK }
\date{\today}

\newlength{\figsize}
\setlength{\figsize}{8.6cm}

\begin{abstract}

With recent experiments investigating the optical properties of progressively 
smaller plasmonic particles, quantum effects become increasingly more relevant, 
requiring a microscopic description. Using the density matrix formalism we analyze
the photo-excited few-electron dynamics of a small plasmonic nanosphere. Following 
the standard derivation of the bulk plasmon we particularly aim on elucidating 
the role of the Coulomb interaction. Calculating the dielectric susceptibility 
spectrum in the linear optical response we find discrete resonances resulting 
from a collective response mediated by the Coulomb interaction between the electrons. 
In the nonlinear regime, the occupations of the system exhibit oscillations between 
the interacting eigenstates.  Our approach provides an ideal platform to study 
and explain nonlinear and quantum plasmonics, revealing that the photo-excited 
dynamics of plasmonic nanospheres has similarities with and combines characteristics 
of both, the well-known two-level Rabi dynamics, and the collective many-electron behavior typical of plasmons. 

\end{abstract}


\maketitle

\section{Introduction}
Quantum plasmonics is a rapidly growing field studying the interaction of light with extremely small metallic systems where quantum effects come into play \cite{tame2013quantum,varas2016quantum,fitzgerald2016quantum}. 
Besides investigating promising applications in chemistry \cite{hartland2017s}, photovoltaics \cite{solar_cell} or photodetection devices \cite{review_Nordlander}, 
an increasing amount of studies are dedicated to more fundamental topics such as the 
emergence of plasmons in nano-systems \cite{varas2016quantum} or the quantization of these plasmonic excitations \cite{tame2013quantum, sundararaman2014theoretical, ballester2009long_range}.
In the simplest picture the plasmon is the optically induced collective oscillation of the 
interacting electron gas against the positive background charge. In the bulk or high density limit a derivation of the plasmon on the microscopic level can be performed using density matrix theory 
\cite{grosso2000solid,giuliani2008quantum,old_paper,LP_in_RPA,dasgupta1977surface,ekardt1984dynamical}. In this derivation the plasmon emerges due to Coulomb interaction between the electrons, 
treated in Hartree-Fock and random phase 
approximation (RPA). When the system size reaches the nanoscale the energy becomes quantized, which suggests the question: What is the impact of a discrete energy structure on the formation of 
the plasmon? 

To gain insight into this fundamental question, we study the dynamics of a few-electron system obtained via optical excitation of a small metal nanosphere. Experimentally, gold nanospheres can 
nowadays be fabricated with sizes less than a few nm \cite{jin2015atomically,weissker2014information} and give rise to interesting nonlinear optical properties \cite{philip2012evolution,knoppe2015nonlinear, knoppe2016second}. 
Current state-of-the-art techniques discussing the emergence of plasmons in nano-systems involve \textit{ab initio} methods, where the optical response of the system is calculated using time-dependent 
density functional theory (TDDFT). Most of the calculations deal with few-atom systems such as one-dimensional sodium chains \cite{yan2007end, yasuike2011collectivity,bernadotte2013plasmons} or 
small metal clusters \cite{ma2015interplay,Govorov_review, Gov}. To distinguish collective, plasmonic excitations from single particle electronic contributions in discrete systems is a challenge 
and there are several studies trying to address the distinction between the two \cite{ma2015interplay,bursi2016quantifying,scholl2012quantum,Zhang2017}. Other studies are trying to bridge the gap between macro- and microscopic 
scales by adapting classical models including electromagnetic calculations \cite{sinha2017classical} and a hydrodynamic model \cite{ciraci2016quantum,ciraci2017current} with the help of correction terms
to take into account quantum effects such as electron spill-out. 

In this work, we follow a different approach to access the optical response of a few-electron system. Similar to what has been done for the bulk case, we use the density matrix formalism and include 
the Coulomb interaction in Hartree-Fock approximation. In the linear regime, we study the change in 
the spectral response of the system. In the non-linear regime, we 
analyze the dynamics of electrons excited at the resonances, which gives rise to an oscillatory behavior strongly influenced by the Coulomb interaction. In a 
diagonalized basis of the mixed states we find an oscillatory dynamics, which we compare to Rabi oscillations found in a two-level system. It turns out that the Coulomb interaction leads to strong 
deviations from two-level system behavior. By explicitly treating the Coulomb terms our paper gives 
valuable insight into the role of Coulomb interaction which is mostly veiled in other approaches.

The paper is organized as follows: the theoretical model is summarized in \cref{sec:theory}, introducing the model Hamiltonian, the analytical wave functions and the derivation of the density matrix
equations of motion for a nanosphere; \cref{sec:results} presents results for a nine-electron model system for two different excitation regimes. The spectral response in the linear regime is discussed
in \cref{subsec:linear}, while the dynamics in the non-linear regime is explored in \cref{subsec:nonlinear}.

\section{Theory\label{sec:theory}}

The Hamiltonian describing the system has three components
\begin{equation}
  \Hop = \Hop_{0}  + \Hop_{ext} + \Hop_{e-e}, \label{eq:H}
\end{equation} 
with the single-electron Hamiltonian $\Hop_0$, the interaction Hamiltonian with an external field $\Hop_{ext}$ and the electron-electron interaction Hamiltonian $\Hop_{e-e}$. 
The single-electron term depends on single-electron eigenenergies $\epsilon_i$ and is given by
\begin{equation}
 \Hop_0 = \sum_{i} \epsilon^{}_{i} \copd_{i} \cop_{i}.
\end{equation}
$\copd_{i}$ and $\cop_{i}$ are the creation and annihilation operators of an electron in a quantum state ${i}$. 
We assume that the electrons are inside the sphere with radius $a$ described by a hard wall potential fulfilling the stationary Schr{\"o}dinger equation. 
The solutions are given by the analytic wave functions $\psi_{i}$ with $i=(NLM)$ in spherical coordinates $\vec{r}=(r,\theta,\phi)$ with the corresponding eigenenergies 
\begin{subequations}
\label{allequations}
\begin{eqnarray} 
\hskip-1.5ex\psi_{NLM}(\vec{r})&=&\sqrt{\dfrac{2}{a^3}}\dfrac{1}{|j_{L+1}(x_{NL})|} j_L\left(x_{NL}\dfrac{r}{a}\right)Y_L^M(\theta,\phi)
\label{eq:wave_func} \\
   \epsilon_{NL} &=& \dfrac{\hbar^2}{2m}\left(\dfrac{x_{NL}}{a}\right)^2, \label{eq:eigenenergies}
\end{eqnarray}
\end{subequations}
where $x_{NL}$ is the $N\textrm{th}$ zero of the spherical Bessel function $j_L$ of order $L$ and $Y_{L}^M$ are the spherical harmonics. The generic set of quantum numbers $i=(NLM)$ 
is adopted from atomic physics and the states are classified as $s,p,d...$-shell, accordingly.
 We do not consider spin in our model. Note that the eigenenergies are $(2L+1)$-fold degenerate.

For the interaction with the external light-field, we consider an electric field polarized in z-direction and we assume dipole approximation (i.e. the field is uniformly distributed 
over the volume of the nanoparticle) 
\begin{equation}
 \Hop_{ext} = -eE(t)\sum_{ij}d^{}_{ij}\copd_{i}\cop_{j},
\end{equation}
where $d_{ij}$ is the dipole matrix element between the states $i$ and $j$. Note that we do not apply the rotating wave approximation (RWA).

\begin{figure}
  \includegraphics[width=\columnwidth]{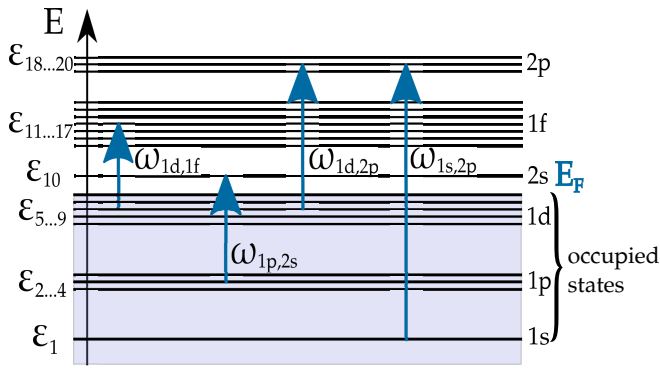}
  \caption{Schematic representation of the allowed dipole transitions in a system with $9$ electrons in $20$ possible interacting states. The energy levels are drawn to scale with degenerate 
  levels split artificially for clarity.\label{fig:en9}}
\end{figure}

The electron-electron interaction is mediated by the Coulomb potential 
\begin{equation}
 \Hop_{e-e} = \dfrac{1}{2}\sum_{ijkl} V^{ijkl} \copd_{i}\copd_{j}\cop_{k}\cop_{l},
\end{equation}
where $V^{ijkl}$ are the Coulomb matrix elements. For the electronic states in \cref{eq:wave_func}, we can calculate 
the Coulomb matrix elements by expanding the potential $\frac{1}{|\mathbf{r}-\mathbf{r'}|}$ using spherical harmonics.
For the interaction with the light field, we assume the quasi-static limit. Assuming that the light field amplitude is
in $z$-direction, the dipole selection rules apply and only states with $\Delta M=0$ and $\Delta L=\pm 1$ can be excited, 
as shown in \cref{fig:en9} for the case of 20 states, i.e. states up to the $2p$-shell.

The dynamics is introduced via the Heisenberg equations of motion 
\begin{equation*}
i\hbar\dfrac{d \hat{\rho}_{nm }}{dt} = [\hat{\rho}_{nm}, \Hop]\,,
\end{equation*}
where $\rho_{nm}=\braket{\hat{\rho}_{nm}}=\braket{\copd_{n}\cop_{m}}$ are the density matrix elements. The diagonal elements of 
the density matrix represent the occupation probability of states in \cref{fig:en9}, while the off-diagonal elements represent 
the coherences between them. The density matrix approach accounts for the time-dependent response of the system. 

Due to its many-body nature, the Coulomb interaction leads to an infinite hierarchy of equations of motion. This is truncated on 
the mean-field (Hartree-Fock) level \cite{Rossi2002}. Hence, only the single particle correlations are taken into account in the carrier-carrier interaction contribution. 
Within this approximation, we obtain the following closed form of the density matrix equations of motion
\begin{align}
 i\hbar\dfrac{d \rho_{nm}}{dt} &= (\epsilon_{m}-\epsilon_n)\rho_{nm} + \nonumber\\ 
		  &\qquad +\sum_{i} \left( W_{m i}\rho_{n i} - W_{i n}\rho_{i m} \right) - \nonumber\\
&\qquad -eE(t)\sum_{i}\left(d_{m i}\rho_{n i}-d_{i n}\rho_{i m} \right),
\label{eq:eom_rho}
\end{align}
where $W(\rho)$ is an effective additional potential induced by the electron-electron interaction, defined as: 
\begin{equation}
 W_{i j} = \sum_{k l} (V^{i k l j}-V^{k i l j})\rho_{kl}. 
 \label{eq:W}
\end{equation}
The first term in \cref{eq:W} is the Hartree term, while the second one is the Fock term which accounts for the particle exchange.
Note that this equation of motion is the starting point for the derivation of the bulk plasmon, if we would take continuous states using $a\to \infty$, as found in standard textbooks \cite{grosso2000solid}.
 
The ground state density matrix of a non-interacting system (i.e. there is no Coulomb interaction between electrons) is diagonal. The electron-electron interaction, on the other hand, 
couples the states and the interacting ground state density matrix will include these coherences.
To obtain the ground state of the interacting system, we slowly turn on the electron-electron interaction by multiplying the Coulomb term in the equations 
of motion \cref{eq:eom_rho} by a switch function with values between $0$ and $1$. We use the following time-dependent switching function:
\begin{equation}
 f(t) = \dfrac{\mathrm{erf}(\alpha(t-t_0))+1}{2},
 \label{eq:switch}
\end{equation}
where the parameter $\alpha$ controls the steepness of the $\textrm{erf}$ function and $t_0$ is the time when the value of $f(t) = 1/2$. 
Using a total simulation time of $2 \ \unit{ps}$ restrains $\alpha \ge 5 \ \unit{ps^{-1}}$ to assure convergence towards
the ground state. This procedure is similar to the adiabatic connection used in Kohn-Sham density functional theory, where 
the non-interacting and interacting electronic systems are explicitly connected by progressively turning on the Coulomb interaction \cite{PhysRevA29, Ernzerhof1996, ad_connection_2014, PhysRevA91}.
Nevertheless, we should point out that the adiabatic connection connects eigenstates of the non-interacting and interacting systems and, 
even if the system is initially in the non-interacting ground state, it may evolve towards an excited state of the interacting Hamiltonian \cite{PhysRevA29}. 
However, we checked that this eigenstate is indeed the ground state by comparing it to the one obtained by minimizing the Hartree-Fock ground state energy. 
Due to the level of approximation applied to the electron-electron interaction term, the two methods are equivalent.

When Coulomb interaction is included, the different states of the density matrix become mixed. We therefore apply a unitary transformation to achieve a ground state in which the density 
matrix is initially diagonal. We call this the interacting basis. The tilde is related to quantities in the interacting basis hereafter, e.g. $\tilde{\rho}$ for the density matrix. 
The equations of motion become
\begin{align}
 i\hbar\dfrac{d \tilde{\rho}_{nm}}{dt} &= \sum_{i} \left(\tilde{\epsilon}_{i m}\tilde{\rho}_{n i}-\tilde{\epsilon}_{n i}\tilde{\rho}_{i m}\right) + \nonumber\\ 
		  &\qquad +\sum_{i} \left( \tilde{W}_{m i}\tilde{\rho}_{n i} - \tilde{W}_{i n}\tilde{\rho}_{i m} \right) - \nonumber\\
&\qquad -eE(t)\sum_{i}\left(\tilde{d}_{m i}\tilde{\rho}_{n i}-\tilde{d}_{i n}\tilde{\rho}_{i m} \right),
\label{eq:eom_rho_ib}
\end{align}
where we defined a matrix $\tilde{\epsilon}$ for the single-electron term as follows
  \begin{equation}
   \tilde{\epsilon}_{ij}= \sum_{k} S^{-1}_{i k} \epsilon_{k} S_{k j}.
   \label{eq:new_eigenen}
  \end{equation}
 $S_{ij}$ are the elements of the transformation matrix. The other two terms in the equations of motion in \cref{eq:eom_rho_ib} are analogous with the ones in the non-interacting basis 
 (\cref{eq:eom_rho}), with new dipole matrix elements
  \begin{equation}
   \tilde{d}_{ij} = \sum_{kl} S_{i k } d_{k l} S^{-1}_{l j}
   \label{eq:new_d}
  \end{equation}
 and new Coulomb matrix elements 
  \begin{equation}
   \tilde{V}^{ijkl} = \sum_{mnop} S_{mi} S_{nj} V^{mnop} S^{-1}_{ok} S^{-1}_{pl}.
   \label{eq:new_V}
  \end{equation}

\section{Results\label{sec:results}}

\subsection{Linear response of a nine-electron system\label{subsec:linear}}
\label{subsec:lin_resp}

We consider a small sphere with radius $a=0.5 \ \unit{nm}$ and we calculate the optical response of a partially filled few-electron system.
Explicitly taking Coulomb interaction into account leads to a rather high number of indices and a complex problem. Hence, we here consider the case of 20 states, 
i.e. we take into account states up to $2p$-shell (see \cref{fig:en9}). Initially, the system is filled with nine electrons up to the $1d$ subshell (spin is not considered). 
The system is excited at an onset time $\mu = 0.2 \, \unit{ps}$ with a Gaussian pulse $E(t) = E_0\exp\left(-\dfrac{(t-\mu)^2}{2\tau_0^2}\right)\sin(\omega_L t)$ with 
$\omega_L = 5 \times 10^{15} \, \unit{s}^{-1}$ (this is an equivalent energy  of $\hbar\omega_0 = 3.29 \, \unit{eV}$), width $\tau_0=0.25 \, \unit{fs}$ and amplitude 
$E_0 = 5.00 \, \unit{mV/m}$. The intensity is sufficiently low such that we are in the linear regime. We then evaluate the optical response by calculating the total 
induced macroscopic polarization $P(t)$ from the microscopic polarizations in the density matrix 
\begin{equation}
 P(t) = \dfrac{e}{v}\sum_{n,m,n<m} 2 \mathscr{Re}(\rho_{nm})d_{nm}  
 \label{eq:Pt}
\end{equation}
where $v=\frac{4\pi a^3}{3}$ denotes the volume of the nanosphere and $d_{nm}$ is the dipole matrix element. 
Using \cref{eq:Pt}, we can calculate the linear response in the frequency domain \cite{nano-optics} 
\begin{equation}
 P(\omega) =\epsilon_0 \chi(\omega) E(\omega). \label{eq:P_omega}
\end{equation}

\begin{figure}
  \includegraphics[width=\columnwidth]{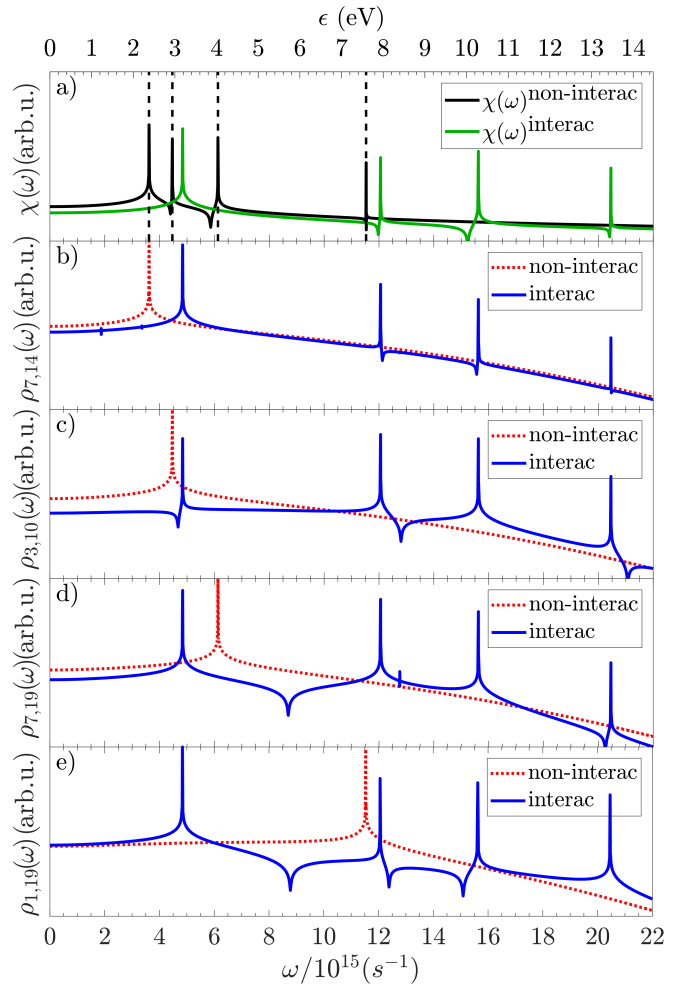}
  \caption{(a) Total induced dielectric susceptibility $\chi(\omega)$ as a function of frequency $\omega$ and energy $\epsilon$ in a system with 9 electrons in 20 states
  with (green) and without (black) Coulomb interaction.(b,c,d, e) The Fourier transform of the light-induced coherences (b) $\rho_{7,14}$ , (c) $\rho_{3,10}$, (d) $\rho_{7,19}$
  and (e) $\rho_{1,19}$ with (blue) and without (dotted red) Coulomb interaction. The values on the y-axis are on logarithmic scale.\label{fig:lin_resp}}
\end{figure}

The linear spectrum for the two cases, non-interacting, i.e. when there is no Coulomb interaction, and interacting, i.e. when Coulomb interaction is present between electrons, 
is shown in \cref{fig:lin_resp}(a). The values displayed on the y-axis are on a logarithmic scale. In the non-interacting case, the spectrum (black curve) contains transitions 
between an empty and a filled state. These occur at the frequencies $\omega_{1d,1f}$, $\omega_{1p,2s}$, $\omega_{1d,2p}$ and $\omega_{1s,2p}$, as also indicated in \cref{fig:en9}. 
The spectrum in \cref{fig:lin_resp} shows discrete resonances at exactly the aforementioned frequencies. The resonances have been labeled and added with dashed lines to \cref{fig:lin_resp}(a). 
The strength of the peaks corresponds to the magnitude of the dipole matrix element $d_{nm}$ and the main peak is the transition between the highest occupied level and 
lowest unoccupied one, i.e. $\omega_{1d,1f}$. This is a transition involving multiple states, where the five-fold degenerate $1d$ shell interacts with five out of the
seven states of $1f$ with $m=\pm2,\pm1 \, \textrm{and} \, 0$, according to the dipole selection rules. Thus, the first resonance is composed of the transitions 
$7 \rightarrow 14$ (corresponding to $\omega_{1d}^{m=0} \rightarrow \omega_{1f}^{m=0}$), $6 \rightarrow 13$ and $8 \rightarrow 15$ ($\omega_{1d}^{m=\pm1}\rightarrow \omega_{1f}^{m=\pm1}$),
$5 \rightarrow 12$ and $9 \rightarrow 16$ ($\omega_{1d}^{m=\pm 2}\rightarrow \omega_{1f}^{m=\pm 2}$, respectively). Similarly for the third resonance, which involves three states. 
For simplicity, we will only discuss the states with $m=0$ in our further analysis, but the behavior is consistent for all degenerate states.    

When the electron-electron interaction is included, the spectrum is blue-shifted (\cref{fig:lin_resp}(a), green curve). Again we find three resonances at $\omega_{1st} = 4.84\times 10^{15} \, 
\unit{s}^{-1}$, $\omega_{2nd} = 12.06 \times 10^{15} \, \unit{s}^{-1}$, $\omega_{3rd} = 15.63 \times 10^{15} \, \unit{s}^{-1}$ and $\omega_{4th} = 20.46 \times 10^{15} \, \unit{s}^{-1}$. 
The discrete resonances cannot be associated with transitions 
between eigenstates of the interacting multi-level system, unlike in the non-interacting case. This is shown by looking at the coherences induced between states by the external light-field in
the frequency domain. As an example, \cref{fig:lin_resp}(b) shows the spectrum of the polarization $\rho_{7,14}(\omega)$. For the non-interacting case it corresponds, as expected, to the peak at 
lowest energy, confirming that here we, indeed, see the transition between those two states. For the interacting case, we find four peaks at the shifted energies. Now, two effects come into play:
first, the Coulomb interaction renormalizes the transition energies between the states leading to a shift, and second and even more interesting, it mixes the response of the individual states, 
indicating the formation of a collective oscillation. Therefore, we conclude that, in the interacting system, the components of the response spectrum cannot be individually assigned to any induced
coherence. Each coherence contributes to the whole spectrum. This is observed for the other three coherences as well, as shown in \cref{fig:lin_resp} (c), (d) and (e). 
We stress out that also in the interacting basis all four coherences ($\tilde{\rho}_{7,14}$, $\tilde{\rho}_{3,10}$, $\tilde{\rho}_{7,19}$ and $\tilde{\rho}_{1,19}$) contribute to each peak.

\begin{figure}
  \includegraphics[width=\columnwidth]{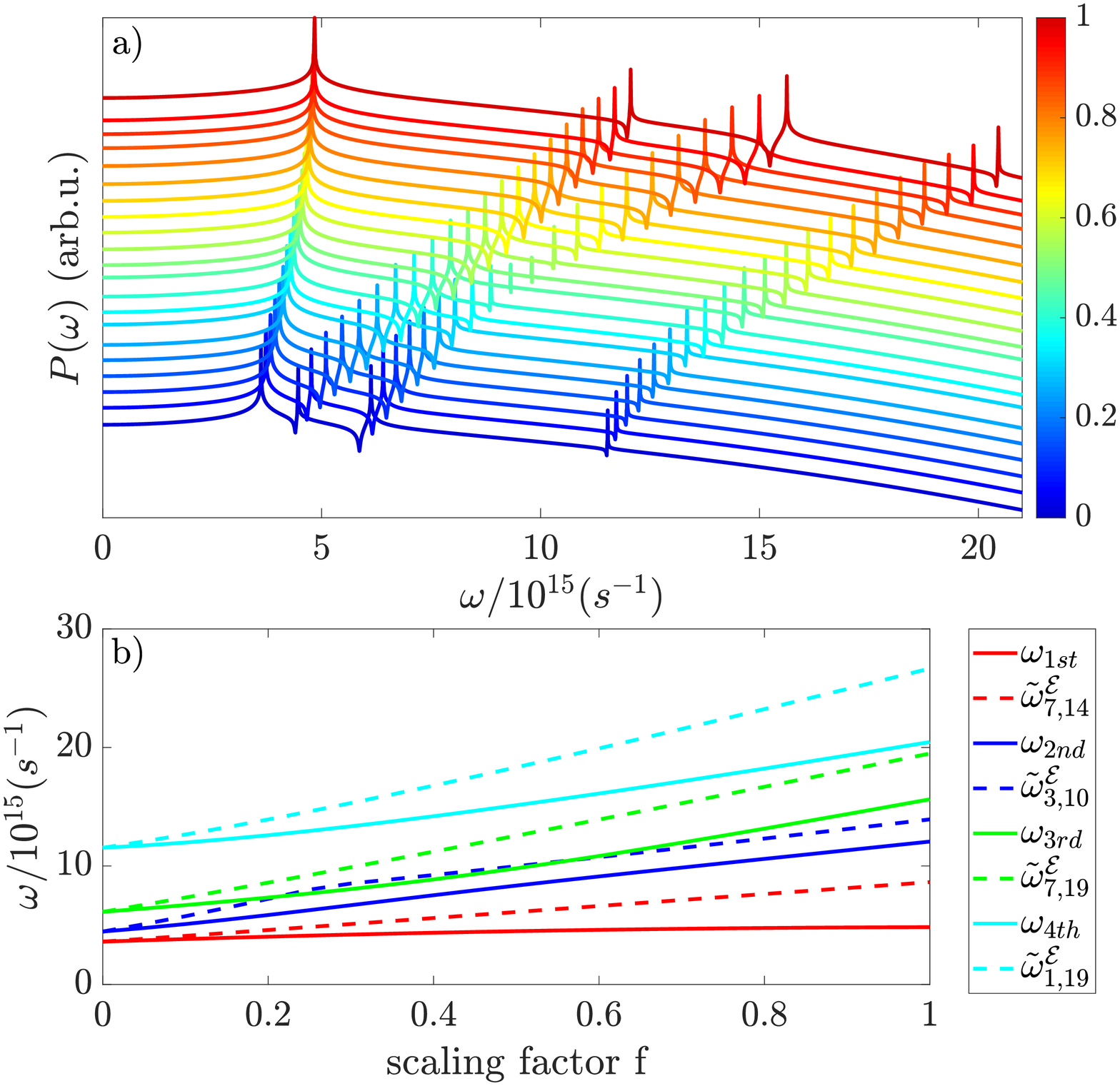} 
  \caption{(a) Shifting of the total induced macroscopic polarization $P(\omega)$ as a function of the scaling factor $f$ of the Coulomb interaction in a system with 9 electrons in 20 states.
  (b) Comparison of the peak positions in the optical response (solid lines) and the eigenvalue difference of the interacting Hamiltonian (dashed lines) as a function of the scaling factor $f$ of 
  the Coulomb interaction. \label{fig:Coulomb_scaling}}
\end{figure}

In order to further investigate the influence of the Coulomb interaction, we introduce a scaling factor $f \in \left[ 0,1 \right]$ multiplied to the Coulomb coupling strength, 
i.e. for $f=0$ we are in the non-interacting case and for $f=1$ in the fully interacting case. Figure \ref{fig:Coulomb_scaling}(a) shows the spectrum for different $f$. 
We can see that the four peaks blue-shift for increasing $f$, i.e., for increasing Coulomb interaction strength. The shift is smooth and derives from the non-interacting peaks. 
The peak position as function of scaling factor $f$ is extracted in Fig.~\ref{fig:Coulomb_scaling}(b) (solid lines). We make two observations: for each of the four peaks the shift
occurs with a different slope and the shift is not strictly linear. 

For a better understanding, we compare these findings to the eigenvalues ${\cal E}$ of the interacting Hamiltonian $H_I = H_0 + W(\rho^{(0)})$ with $\rho^{(0)}$ being the density
matrix of the Coulomb correlated ground state. Note that the eigenvalues ${\cal{E}}$ cannot be directly compared to the matrix elements of $\tilde{\epsilon}$ obtained by diagonalizing
the interacting initial state defined in \cref{sec:theory}. If the Coulomb interaction only leads to energy renormalization we would expect that the peak positions of the spectrum agree 
with the energy difference of the eigenvalues $\hbar\omega^{\cal{E}}_{ij}={\cal{E}}_{j}-{\cal{E}}_{i}$. Seeing that the Coulomb correlated state can be traced back to the non-interacting 
states, we consider the states corresponding to the dipole transitions $\omega^{\cal{E}}_{7,14}$, $\omega^{\cal{E}}_{3,10}$, $\omega^{\cal{E}}_{7,19}$ and $\omega^{\cal{E}}_{1,19}$. 

The comparison between the peak position (solid lines) and the eigenvalue difference (dashed lines) is shown in Fig.~\ref{fig:Coulomb_scaling}(b). 
Naturally for the non-interacting case of $f=0$ the two curves agree. 
But for $f\neq 0$ we find a strong disagreement between the two curves, where the peak positions are always below the eigenvalue difference. 
A noticeable feature occurs for a scaling factor $f\approx 0.3$, where the second and third peaks in the polarization spectrum show an anti-crossing in \cref{fig:Coulomb_scaling}(a).
This is due to the fact that the Coulomb interaction changes the order of the $2s$ and $1f$ orbitals as it can be verified by checking the order of the eigenvalues ${\cal E}$ of the interacting Hamiltonian $H_I$.
Indeed, around $f=0.3$, the blue dashed curve $\omega^{\cal{E}}_{3,10}$ in \cref{fig:Coulomb_scaling}(b) bents, indicating where the swap occurs.  
This mismatch between the eigenenergies and the polarization spectra is reminiscent of excitons in semiconductors whose spectral response always lies below the band gap. 
Excitons are a many-particle phenomena, where the Coulomb interaction correlates the 
individual electron energies to form a collective, bound state. This indicates the importance of the non-linear effects induced by the Coulomb interaction as well as the state mixing. It is also 
a hint for collectivity in the system, because all states contribute to the formation of the peaks.

\subsection{Non-linear response of a nine-electron system\label{subsec:nonlinear}}

In this section, we analyze the optical response of the system beyond the linear regime. We calculate the density matrix time-evolution for a strong excitation at each of the four resonances in the 
interacting polarization spectrum (\cref{fig:lin_resp}a)) under a continuous wave excitation that is switched on instantaneously. Accordingly we are interested in the response to excitations at
$\omega_{1st} = 4.84 \times 10^{15} \ \unit{s}^{-1}$, $\omega_{2nd} = 12.06 \times 10^{15} \ \unit{s}^{-1}$, $\omega_{3rd} = 15.63 \times 10^{15} \ \unit{s}^{-1}$ and $\omega_{4th} = 20.46 
\times 10^{15} \, \unit{s}^{-1}$. The corresponding electric field reads $E(t) = E_0\sin(\omega_L t)\Theta(t)$ with $E_0$ being the amplitude, $\omega_L$ the frequency of the field and 
$\Theta(t)$ the Heaviside function. 

\begin{figure}[t]
  \includegraphics[width=\columnwidth]{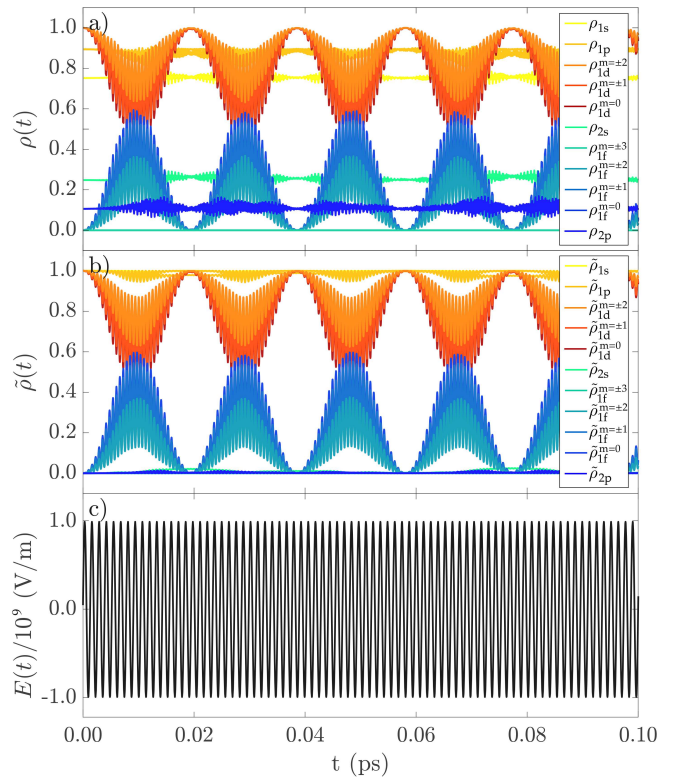}
  \caption{Population dynamics in (a) the non-interacting and (b) the interacting basis excited by a pulse with amplitude $E_0=10^9 \ \unit{V/m}$ and frequency 
  $\omega_L=\omega_{1st}$ with (c) the corresponding electric field.\label{fig:pop_dyn}}
\end{figure}

We start by analyzing an excitation with frequency close to the first peak in the absorption spectrum $\omega_{1st}$. In the non-interacting basis due to the adiabatic switch-on into the Coulomb
correlated basis, not only the lowest nine states are occupied, but most states have a finite initial occupation. When the electric field sets in, an oscillatory dynamics is initiated, as can be
seen in \cref{fig:pop_dyn}(a), induced by a light field with a frequency at $\omega_L=\omega_{1st}$ and an amplitude $E_0=10^9 \ \unit{V/m}$ displayed in \cref{fig:pop_dyn}(c). 
The strongest oscillation occurs between the states in $1d$ and $1f$ and, similarly with the non-interacting case, it is the main contributor to the first peak in the linear response spectrum. 
The superimposed fast oscillation can be traced back to the fact that we do not perform the RWA. Additionally all the other states start to oscillate showing that they contribute 
to the dynamics in a non-trivial way. In the non-interacting basis, we can still attribute the oscillation of all states to the state mixing in the initial state. 
This is in agreement with our finding in the linear regime, where we also found that several states contribute to a single peak.

Now we switch to the interacting basis, where the initial state is diagonal. This allows us to study the question, whether we can describe the system by an excitation between only two Coulomb-correlated states. 
Figure~\ref{fig:pop_dyn}(b) shows the population dynamics of the density matrix in the interacting basis $\tilde{\rho}(t)$. We find that the population exchange occurs mainly between the $1d$ and $1f$
interacting states, $\tilde{\rho}_{1d}^{m=\pm2,\pm1,0}$ and $\tilde{\rho}_{1f}^{m=\pm2,\pm1,0}$ (red and blue curves in \cref{fig:pop_dyn}(b)), again because $\tilde{\rho}_{1d,1f}$ is the strongest dipole-allowed 
transition in the system.
We also observe oscillations with a much lower amplitude of other states like $\tilde{\rho}_{1p}$, $\tilde{\rho}_{1s}$ and $\tilde{\rho}_{2p}$, which also have a dipole matrix element between them. 

\begin{figure}[t]
  \includegraphics[width=\columnwidth]{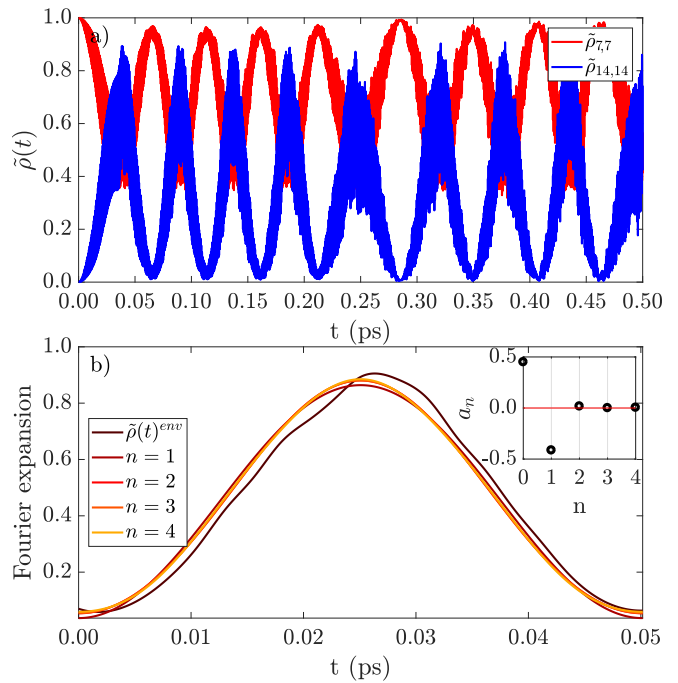}
  \caption{a) Single oscillation of the occupations $\tilde{\rho}_{7,7}$ and $\tilde{\rho}_{14,14}$ for an excitation frequency $\omega_L = 5.07 \times 10^{15} \ \unit{s}^{-1}$ and 
  $E_0=3.4\times 10^{8}\ \unit{V/m}$ ($\Omega^{7,14}/\omega_{1st} = 0.0178$).
  b) Comparison with the Fourier expansion of the occupation $\tilde{\rho}_{14,14}$ up to order $n$ with the cosine coefficients $a_n$ as inset.
  \label{fig:F_comp1}}
\end{figure}

An oscillation amplitude of $\approx 1$ between the states $\tilde{\rho}_{7,7}$ and $\tilde{\rho}_{14,14}$ can be achieved by slightly changing the excitation frequency to $\omega_L=5.07\times 10^{15} 
\, \unit{s}^{-1}> \omega_{1st}$, which is slightly higher than the resonance frequency found in the linear regime. 
Nevertheless, even for this excitation condition, there is no perfect inversion of the system between $0$ and $1$ for the occupation of $\tilde{\rho}_{14,14}$. This is because of the many-body character of
the electronic system where the population transfer occurs between multiple states, mediated by either a dipole or a Coulomb matrix element. 
The occupation approximately follows a pure sine (or cosine) behavior. We confirm these findings by a Fourier analysis of the oscillation. We consider the Fourier 
expansion $\tilde{\rho}(t) = a_0+\sum_n a_n\cos\left(n\Omega_0 t\right)$ of the oscillation. Only $a_0=0.5$ and $a_1=0.5$ are different from $0$, resulting in the Fourier function
$ = 0.5+ 0.5 \cos\left(n\Omega_0 t\right)$ with $\Omega_0=1.26\times 10^{14} \, \unit{s^{-1}}$. This is displayed in Fig.~\ref{fig:F_comp1}, where we show the Fourier expansion up to order $n=4$ and compare it to a
single oscillation of the dynamics. 
In order to get rid of the small oscillations which are present because we have not performed the RWA and because of the degeneracy of the states involved in the interaction,
we extract the envelope of one of the cycles of the time dynamics.

\begin{figure}[t]
 \includegraphics[width=\columnwidth]{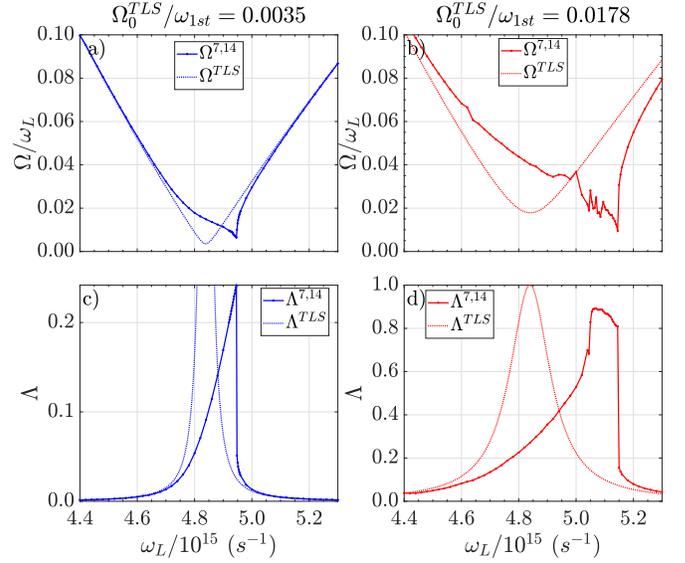}
  \caption{a), b) Frequency $\Omega$ and c), d) amplitude $\Lambda$ for excitations close to the first peak $\omega_{1st}$ for an excitation with $\Omega^{7,14}/\omega_{1st} = 0.0035$ (left, blue curves)
  and $\Omega^{7,14}/\omega_{1st} = 0.0178$ (right, red curves). The solid lines are the extracted values from the dynamics $\Omega^{7,14}$ and $\Lambda^{7,14}$ and the dotted lines are from the TLS 
  $\Omega^{TLS}$ and $\Lambda^{TLS}$.\label{fig:RO_1st_peak}}
\end{figure}

To analyze the oscillations in a more systematic way, we sweep the excitation frequency of the light field through the resonance. From the time-series of $\tilde{\rho}_{14,14}$ (note that $\tilde{\rho}_{7,7}$
behaves analogous, as well as all the other states in $1d$ and $1f$) we extract the frequency $\Omega^{7,14}=\dfrac{2\pi}{\tau^{7,14}}$ with $\tau^{7,14}$ being the period as well as the amplitude 
$\Lambda^{7,14}=\textrm{max}(\tilde{\rho}_{14,14})$ of the oscillation for two different excitation strengths $\Omega^{7,14}/\omega_{1st} = 0.0035$ (corresponding to a light-field amplitude 
$E_0=6.8 \times 10^7 \, \unit{V/m}$) shown in Fig.~\ref{fig:RO_1st_peak} (left column) and 
$\Omega^{7,14}/\omega_{1st} = 0.0178$ ($E_0=3.4 \times 10^8 \, \unit{V/m}$) shown in Fig.~\ref{fig:RO_1st_peak} (right column). The ratio between the Rabi frequency and the light frequency gives 
a relative measure of how excited the system is at a certain light frequency and intensity, beyond the absolute value of the field strength and the dipole moment of the transition. In this way, 
we can compare different resonances in the light-induced polarization at approximately the same excitation conditions. 
For the stronger excitation regime (right column in Fig.~\ref{fig:RO_1st_peak}) we see a resonant behavior with a minimal frequency $\Omega^{7,14}$ and an amplitude of $\Lambda^{7,14}=0.9$ 
occurring at $\omega_{1st}=5.07 \times 10^{15}$, which is slightly higher than the resonance frequency found in the linear regime. Similarly for the weaker excitation (left column in Fig.~\ref{fig:RO_1st_peak}), 
where the resonance shifts to $\omega_{1st}=4.94 \times 10^{15}$. Nevertheless, the state is only partially occupied $\Lambda^{7,14}=0.25$. 

The oscillations in \cref{fig:pop_dyn}b) indicate that the population dynamics of the interacting states behaves similarly to a two-level system (TLS) showing Rabi oscillations. 
For a resonant excitation, the Rabi frequency in a TLS is defined as $\Omega_0 = e|\tilde{d}_{i j}|E_0/\hbar$ with $\tilde{d}_{ij}$ the dipole matrix element in the interacting basis (cf. \cref{eq:new_d}). 
The generalized Rabi frequency is then
\begin{equation}
 \Omega^{TLS} = \sqrt{(\Omega_0^{TLS})^2 + \Delta^2},
 \label{eq:Rabi_TLS}
\end{equation}
where $\Delta = \omega - \omega_0$ is the detuning from the resonant frequency $\omega_0$. Remember, that only for $\Delta=0$ the oscillation amplitude in a TLS is $\Lambda^{TLS}=1$.
For off-resonant excitation the amplitude $\Lambda^{TLS}$ is given by
\begin{equation}
 \Lambda^{TLS} = \dfrac{\Omega_0^2}{\Omega_0^2+\Delta^2}\,.
 \label{eq:Rabi_amplitude}
\end{equation}

We now compare our findings of the frequency sweep to the Rabi oscillations in a TLS. For the TLS we take the peak position at $\omega_0=\omega_{1st}= 4.84 \times 10^{15} \, \unit{s}^{-1}$
as resonance frequency. We start with the right column in Fig.~\ref{fig:RO_1st_peak}. For this large excitation strength the oscillations of the occupations become faster.
This leads to an effective higher occupation of the previously unoccupied states 
and in return modifies the Coulomb interaction resulting in a shifted curve and the observed differences. As opposed to a simple two-level system, there is an unstable excitation window where the occupation
is large and the Rabi frequency is fluctuating, followed by a discontinuous drop in amplitude (increase in frequency). 
Thus, the Coulomb interaction influences the dynamics of the system in a complex way.

The weaker excitation $\Omega^{7,14}/\omega_{1st} = 0.0035$ (see \cref{fig:RO_1st_peak}a) and c)) behaves less pronounced. For a large range of values of excitation frequency $\omega$ away from resonance
we find an excellent agreement between the extracted values and the prediction via the TLS. However, close to resonance these two pictures deviate strongly. The minimal oscillation frequency $\Omega$
and the maximal amplitude $\Lambda$ is again shifted to higher excitation frequencies compared to the predictions of the TLS $\Omega^{TLS}$ and $\Lambda^{TLS}$. 
Remarkably, the amplitude always stays far below 1 with the highest $\Lambda^{7,14}=0.25$.
We attribute this to the change in occupations
leading to a change in the Coulomb interaction which dynamically shifts the resonance frequency. Note also that the slope of the two curves are different: while the TLS exhibits a symmetric behavior
with respect to the resonance, the extracted values show a kink. This again hints towards non-linear and collective effects induced by the Coulomb interaction that is not simply a shift in resonance frequency.

\begin{figure}[t]
\includegraphics[width=\columnwidth]{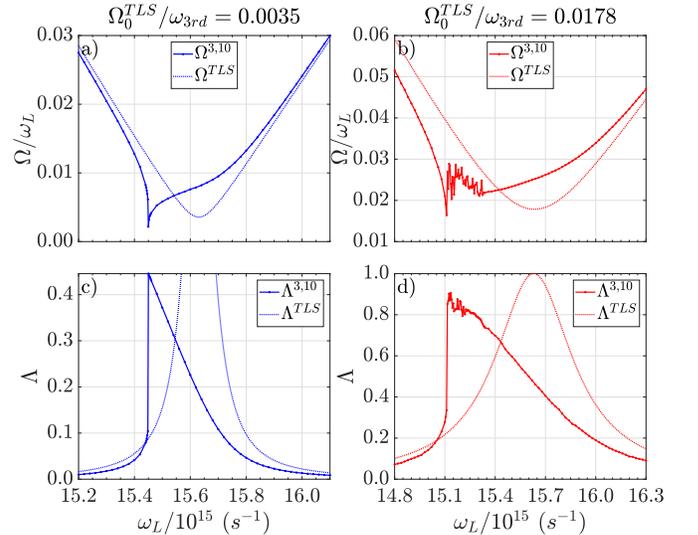}
  \caption{Same as \cref{fig:RO_1st_peak}, but for excitations close to the third peaks $\omega_{3rd}$. We consider  $\Omega^{3,10}$ and $\Lambda^{3,10}$.  \label{fig:RO_3rd_peak}}
\end{figure}

Next, we want to consider what happens if we excite the system with frequencies close to the third peak at $\omega_{3rd}=15.63\times 10^{15} \ \unit{s}^{-1}$. We remark that the findings are similar to what
happens at the second and fourth peak which we therefore do not show. Again, we find an oscillatory behavior in the occupation of the states, this time the main oscillation in the 
interacting basis happens between $\tilde{\rho}_{3,3}$ and $\tilde{\rho}_{10,10}$ in agreement with the dipole moment $\tilde{d}_{1p,2s}$ corresponding to this transition. Although 
this transition contributes to the second peak in the $\chi(\omega)$ spectrum in the case without Coulomb interaction, it becomes the main contributor to the third peak in the case 
with Coulomb interaction. This is due to the change in the order of the $2s$ and $1f$ orbitals, as explained at the end of \cref{subsec:lin_resp}.  

Therefore we do a systematic analysis of the frequency $\Omega^{3,10}$ and amplitude $\Lambda^{3,10}$ corresponding to the dynamics of the population of $\tilde{\rho}_{10,10}$. 
The corresponding results are shown in Fig.~\ref{fig:RO_3rd_peak} for two different pulse strengths such that the ratios $\Omega^{3,10}/\omega_{3rd}$ are the same as for the first peak.
We again compare our findings to the behavior of a TLS, here taken with a resonance frequency of $\omega_0=\omega_{3rd}$.

In Fig.~\ref{fig:RO_3rd_peak} (right column) we consider an excitation with a light field $E_0 = 5 \times 10^9 \, \unit{V/m}$ inducing a ratio $\Omega^{3,10}/\omega_{3rd}=0.0178$. 
Again, there is a strong discrepancy between our system and the theoretical two-level system. There is almost total population inversion for an extended excitation window, with an asymmetric 
and discontinuous excursion amplitude. This time, the kink is towards lower frequencies.

For the weaker regime, $\Omega^{3,10}/\omega_{3rd}=0.0035$ and the corresponding absolute field strength $E_0 = 10^9 \, \unit{V/m}$, far off the resonance the extracted values and the TLS prediction agree well,
while close to the resonance we see a kink in the frequency behavior of $\Omega^{3,10}$. The amplitude always stays far below 1 with the highest $\Lambda^{3,10}=0.45$.

\begin{figure}[t]
  \includegraphics[width=\columnwidth]{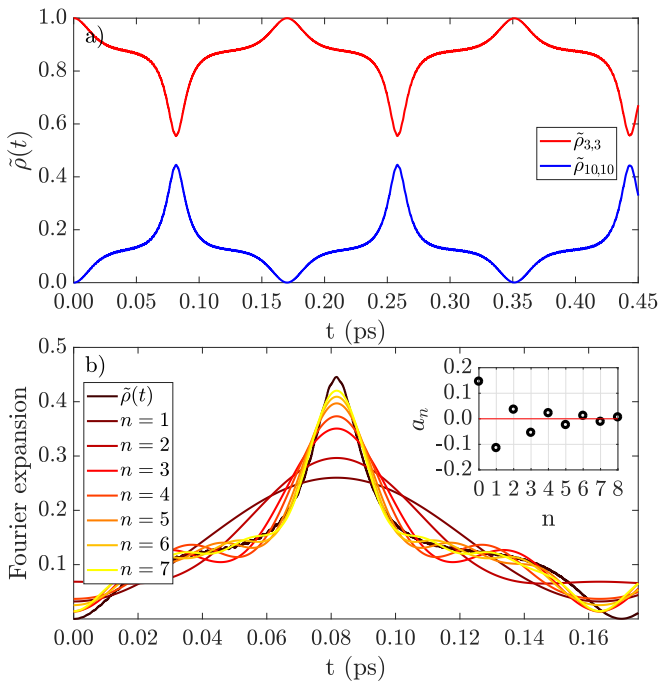}
  \caption{a) Population dynamics of $\tilde{\rho}_{3,3}$ and $\tilde{\rho}_{10,10}$ for an excitation at $\omega_L = 15.45 \times 10^{15} \ \unit{s}^{-1}$
  and  $E_0 = 10^9 \, \unit{V/m}$ ($\Omega^{3,10}/\omega_{3rd}=0.0035$). 
  b) Fourier series $\tilde{\rho}_{10,10}$, where each curve $a_n$ implies that cosine coefficients up to order $n$ have been added to the Fourier series with the values of the cosine coefficients $a_n$ as inset.
 \label{fig:F_comp3}}
\end{figure}

Finally, let us have a closer look at the population dynamics of the contributing states $\tilde{\rho}_{3,3}$ and $\tilde{\rho}_{10,10}$ in the interacting basis. 
Two periods of the oscillation are shown in \cref{fig:F_comp3} induced by a field at the maximal amplitude using $\omega_L = 15.45 \times 10^{15} \ \unit{s}^{-1}$ 
and $\Omega^{3,10}/\omega_{3rd}=0.0035$. The dynamics is similar for $\tilde{\rho}_{7,7}$ and $\tilde{\rho}_{14,14}$ at maximum amplitude and same excitation regime. 
Already here, we clearly see that the dynamical behavior deviates in 
an interesting way from the sine-behavior found in a TLS. It is more bell-shaped.
Similar bell-shaped oscillations have been found in the Rabi oscillations in semiconductor quantum dots under local fields,
which likewise result from the non-linear term in the Bloch equations \cite{Slepyan2004}. When we do a Fourier analysis of
this curve, we find that higher harmonics contribute strongly to the formation of the dynamics, in this case up to sixth
order. This can again be traced back to the Coulomb interaction, which has more influence when the pulse strength is weaker. 
Hence, the influence of the Coulomb interaction leads to non-linear effects like the instantaneous formation of higher
harmonics for the light-induced changes in the occupation. 
Although pumping the system harder reduces the order of the Rabi frequency harmonics induced (cf. \cref{fig:F_comp1}),
the non-linear effects are still manifesting via the asymmetric occupations as a function of the excitation frequency and the
complex time-dynamics around the resonance.   

We emphasize that we distinguish two types of non-linear effects in our system: higher harmonics of the Rabi oscillations' frequency and dynamical shift
of the resonance frequency. The two effects are competing with each other since both the Coulomb interaction and the light-field are driving the dynamics via the Hamiltonian. 
Hence, for weaker excitation, the Coulomb interaction dominates, resulting in the bell-shaped dynamics, while a stronger one 
drives the oscillations closer to an optical two-level system excitation.

To conclude this section, we comment on our method in relation to DFT methods. We remark that our formulation takes into account a bare Coulomb potential calculated 
explicitly from the analytical single-electron states within the Hartree-Fock (exchange) approximation. In contrast, in DFT going 
beyond Hartree approximation involves considering an exchange-correlation
 functional such as the local density approximation (LDA). Within this approximation, a DFT ground state is calculated, from which a linear response can be obtained. 
 Because the ground state from our method and the DFT ground state can differ quite significantly, we expect also quite different 
results comparing the two methods. To calculate the dynamics, it is possible to use (real-time) time-dependent (RT-TDDFT) codes. 
In these calculations, the exchange-correlation functional is again a crucial ingredient, which due to its dynamical shift has to be treated with great care \cite{
ruggenthaler2009rabi,fuks2011nonlinear,provorse2016electron}, while our approach is well suited to study dynamics. We would also like to remark that in DFT 
studies the coherences between the states might be lost due to the reduction to the electronic density, while in our dynamics we naturally keep 
all the coherences between the states.
 
While the TDDFT methods are widely used in the study of quantum plasmonic systems, the density matrix approach is a well-established method in the field
 of optically excited semiconductors, particularly to describe Coulomb effects like excitons and band gap renormalization \cite{Rossi2002,haug2009quantum}. Due
 to the discrete energy levels in our system and the finite gap between the highest occupied and lowest unoccupied state, the system is somewhere in between a 
 metal and a semiconductor. 

\section{Conclusions}

In summary, we have used a microscopic density matrix formalism to describe the electrons dynamics in a quantized nanoparticle. The methodology is similar 
to the one used for the derivation of bulk plasmons. We considered a system of $9$ electrons with an explicit description of the Coulomb 
interaction on the mean field level and studied the effects of the Coulomb interaction on the optically induced dynamics. In the linear regime, 
the Coulomb interaction leads to a shift of the resonances and a state mixing, indicating a collective response. In the non-linear regime, the optically 
induced oscillations of the occupation were strongly modified due to the Coulomb interaction.

Our study lies in between two limiting cases: in one limit, one considers one or two electrons which then results in the description of Rabi oscillations
and a formulation using a two-level system. Such a description is used successfully, for example, to describe semiconductor quantum dots. In the other limit, 
one considers many-electron systems where plasmons are described within the RPA, and a classical treatment of the material polarizability is usually sufficient. 
Neither of the aforementioned descriptions is sufficient to describe the Coulomb effects in a few-electron system, underlined by the fact that our findings 
deviate from both limiting cases. Instead, we find similarities with both cases. For certain excitation conditions the system behaves like the two-level 
system exhibiting Rabi-oscillations. On the other hand, we find hints for collective behavior, like energy shifts and the modification of the oscillations, 
which are typical for plasmons. 

\section*{Acknowledgments}
This work was supported by the EPSRC through EP/L027151/1.
We are grateful to Benjamin Burnett and Stefano Guazzotti for most fruitful discussions. A. Crai thanks Johannes Lischner for the helpful advice. 
D.E.R. is grateful for financial support from the German Academic Exchange Service (DAAD) within the P.R.I.M.E. programme.


\begin{thebibliography}{42}%
\makeatletter
\providecommand \@ifxundefined [1]{%
 \@ifx{#1\undefined}
}%
\providecommand \@ifnum [1]{%
 \ifnum #1\expandafter \@firstoftwo
 \else \expandafter \@secondoftwo
 \fi
}%
\providecommand \@ifx [1]{%
 \ifx #1\expandafter \@firstoftwo
 \else \expandafter \@secondoftwo
 \fi
}%
\providecommand \natexlab [1]{#1}%
\providecommand \enquote  [1]{``#1''}%
\providecommand \bibnamefont  [1]{#1}%
\providecommand \bibfnamefont [1]{#1}%
\providecommand \citenamefont [1]{#1}%
\providecommand \href@noop [0]{\@secondoftwo}%
\providecommand \href [0]{\begingroup \@sanitize@url \@href}%
\providecommand \@href[1]{\@@startlink{#1}\@@href}%
\providecommand \@@href[1]{\endgroup#1\@@endlink}%
\providecommand \@sanitize@url [0]{\catcode `\\12\catcode `\$12\catcode
  `\&12\catcode `\#12\catcode `\^12\catcode `\_12\catcode `\%12\relax}%
\providecommand \@@startlink[1]{}%
\providecommand \@@endlink[0]{}%
\providecommand \url  [0]{\begingroup\@sanitize@url \@url }%
\providecommand \@url [1]{\endgroup\@href {#1}{\urlprefix }}%
\providecommand \urlprefix  [0]{URL }%
\providecommand \Eprint [0]{\href }%
\providecommand \doibase [0]{http://dx.doi.org/}%
\providecommand \selectlanguage [0]{\@gobble}%
\providecommand \bibinfo  [0]{\@secondoftwo}%
\providecommand \bibfield  [0]{\@secondoftwo}%
\providecommand \translation [1]{[#1]}%
\providecommand \BibitemOpen [0]{}%
\providecommand \bibitemStop [0]{}%
\providecommand \bibitemNoStop [0]{.\EOS\space}%
\providecommand \EOS [0]{\spacefactor3000\relax}%
\providecommand \BibitemShut  [1]{\csname bibitem#1\endcsname}%
\let\auto@bib@innerbib\@empty
\bibitem [{\citenamefont {Tame}\ \emph {et~al.}(2013)\citenamefont {Tame},
  \citenamefont {McEnery}, \citenamefont {{\"O}zdemir}, \citenamefont {Lee},
  \citenamefont {Maier},\ and\ \citenamefont {Kim}}]{tame2013quantum}%
  \BibitemOpen
  \bibfield  {author} {\bibinfo {author} {\bibfnamefont {M.}~\bibnamefont
  {Tame}}, \bibinfo {author} {\bibfnamefont {K.}~\bibnamefont {McEnery}},
  \bibinfo {author} {\bibfnamefont {S.}~\bibnamefont {{\"O}zdemir}}, \bibinfo
  {author} {\bibfnamefont {J.}~\bibnamefont {Lee}}, \bibinfo {author}
  {\bibfnamefont {S.}~\bibnamefont {Maier}}, \ and\ \bibinfo {author}
  {\bibfnamefont {M.}~\bibnamefont {Kim}},\ }\href@noop {} {\bibfield
  {journal} {\bibinfo  {journal} {Nat. Phys.}\ }\textbf {\bibinfo {volume}
  {9}},\ \bibinfo {pages} {329} (\bibinfo {year} {2013})}\BibitemShut {NoStop}%
\bibitem [{\citenamefont {Varas}\ \emph {et~al.}(2016)\citenamefont {Varas},
  \citenamefont {Garc{\'\i}a-Gonz{\'a}lez}, \citenamefont {Feist},
  \citenamefont {Garc{\'\i}a-Vidal},\ and\ \citenamefont
  {Rubio}}]{varas2016quantum}%
  \BibitemOpen
  \bibfield  {author} {\bibinfo {author} {\bibfnamefont {A.}~\bibnamefont
  {Varas}}, \bibinfo {author} {\bibfnamefont {P.}~\bibnamefont
  {Garc{\'\i}a-Gonz{\'a}lez}}, \bibinfo {author} {\bibfnamefont
  {J.}~\bibnamefont {Feist}}, \bibinfo {author} {\bibfnamefont
  {F.}~\bibnamefont {Garc{\'\i}a-Vidal}}, \ and\ \bibinfo {author}
  {\bibfnamefont {A.}~\bibnamefont {Rubio}},\ }\href@noop {} {\bibfield
  {journal} {\bibinfo  {journal} {Nanophotonics}\ }\textbf {\bibinfo {volume}
  {5}},\ \bibinfo {pages} {409} (\bibinfo {year} {2016})}\BibitemShut {NoStop}%
\bibitem [{\citenamefont {Fitzgerald}\ \emph {et~al.}(2016)\citenamefont
  {Fitzgerald}, \citenamefont {Narang}, \citenamefont {Craster}, \citenamefont
  {Maier},\ and\ \citenamefont {Giannini}}]{fitzgerald2016quantum}%
  \BibitemOpen
  \bibfield  {author} {\bibinfo {author} {\bibfnamefont {J.~M.}\ \bibnamefont
  {Fitzgerald}}, \bibinfo {author} {\bibfnamefont {P.}~\bibnamefont {Narang}},
  \bibinfo {author} {\bibfnamefont {R.~V.}\ \bibnamefont {Craster}}, \bibinfo
  {author} {\bibfnamefont {S.~A.}\ \bibnamefont {Maier}}, \ and\ \bibinfo
  {author} {\bibfnamefont {V.}~\bibnamefont {Giannini}},\ }\href@noop {}
  {\bibfield  {journal} {\bibinfo  {journal} {Proc. IEEE}\ }\textbf {\bibinfo
  {volume} {104}},\ \bibinfo {pages} {2307} (\bibinfo {year}
  {2016})}\BibitemShut {NoStop}%
\bibitem [{\citenamefont {Hartland}\ \emph {et~al.}(2017)\citenamefont
  {Hartland}, \citenamefont {Besteiro}, \citenamefont {Johns},\ and\
  \citenamefont {Govorov}}]{hartland2017s}%
  \BibitemOpen
  \bibfield  {author} {\bibinfo {author} {\bibfnamefont {G.~V.}\ \bibnamefont
  {Hartland}}, \bibinfo {author} {\bibfnamefont {L.}~\bibnamefont {Besteiro}},
  \bibinfo {author} {\bibfnamefont {P.}~\bibnamefont {Johns}}, \ and\ \bibinfo
  {author} {\bibfnamefont {A.~O.}\ \bibnamefont {Govorov}},\ }\href@noop {}
  {\bibfield  {journal} {\bibinfo  {journal} {ACS Energy Lett.}\ }\textbf
  {\bibinfo {volume} {2}},\ \bibinfo {pages} {1641} (\bibinfo {year}
  {2017})}\BibitemShut {NoStop}%
\bibitem [{\citenamefont {Atwater}\ and\ \citenamefont
  {Polman}(2010)}]{solar_cell}%
  \BibitemOpen
  \bibfield  {author} {\bibinfo {author} {\bibfnamefont {H.~A.}\ \bibnamefont
  {Atwater}}\ and\ \bibinfo {author} {\bibfnamefont {A.}~\bibnamefont
  {Polman}},\ }\href@noop {} {\bibfield  {journal} {\bibinfo  {journal} {Nat.
  Mater.}\ }\textbf {\bibinfo {volume} {9}},\ \bibinfo {pages} {205} (\bibinfo
  {year} {2010})}\BibitemShut {NoStop}%
\bibitem [{\citenamefont {Brongersma}\ \emph {et~al.}(2015)\citenamefont
  {Brongersma}, \citenamefont {Halas},\ and\ \citenamefont
  {Nordlander}}]{review_Nordlander}%
  \BibitemOpen
  \bibfield  {author} {\bibinfo {author} {\bibfnamefont {M.~L.}\ \bibnamefont
  {Brongersma}}, \bibinfo {author} {\bibfnamefont {N.~J.}\ \bibnamefont
  {Halas}}, \ and\ \bibinfo {author} {\bibfnamefont {P.}~\bibnamefont
  {Nordlander}},\ }\href@noop {} {\bibfield  {journal} {\bibinfo  {journal}
  {Nat. Nanotechnol.}\ }\textbf {\bibinfo {volume} {10}},\ \bibinfo {pages}
  {25} (\bibinfo {year} {2015})}\BibitemShut {NoStop}%
\bibitem [{\citenamefont {Sundararaman}\ \emph {et~al.}(2014)\citenamefont
  {Sundararaman}, \citenamefont {Narang}, \citenamefont {Jermyn}, \citenamefont
  {Goddard~III},\ and\ \citenamefont {Atwater}}]{sundararaman2014theoretical}%
  \BibitemOpen
  \bibfield  {author} {\bibinfo {author} {\bibfnamefont {R.}~\bibnamefont
  {Sundararaman}}, \bibinfo {author} {\bibfnamefont {P.}~\bibnamefont
  {Narang}}, \bibinfo {author} {\bibfnamefont {A.~S.}\ \bibnamefont {Jermyn}},
  \bibinfo {author} {\bibfnamefont {W.~A.}\ \bibnamefont {Goddard~III}}, \ and\
  \bibinfo {author} {\bibfnamefont {H.~A.}\ \bibnamefont {Atwater}},\
  }\href@noop {} {\bibfield  {journal} {\bibinfo  {journal} {Nat. Commun.}\
  }\textbf {\bibinfo {volume} {5}},\ \bibinfo {pages} {5788} (\bibinfo {year}
  {2014})}\BibitemShut {NoStop}%
\bibitem [{\citenamefont {Ballester}\ \emph {et~al.}(2009)\citenamefont
  {Ballester}, \citenamefont {Tame}, \citenamefont {Lee}, \citenamefont {Lee},\
  and\ \citenamefont {Kim}}]{ballester2009long_range}%
  \BibitemOpen
  \bibfield  {author} {\bibinfo {author} {\bibfnamefont {D.}~\bibnamefont
  {Ballester}}, \bibinfo {author} {\bibfnamefont {M.~S.}\ \bibnamefont {Tame}},
  \bibinfo {author} {\bibfnamefont {C.}~\bibnamefont {Lee}}, \bibinfo {author}
  {\bibfnamefont {J.}~\bibnamefont {Lee}}, \ and\ \bibinfo {author}
  {\bibfnamefont {M.~S.}\ \bibnamefont {Kim}},\ }\href@noop {} {\bibfield
  {journal} {\bibinfo  {journal} {Phys. Rev. A}\ }\textbf {\bibinfo {volume}
  {79}},\ \bibinfo {pages} {053845} (\bibinfo {year} {2009})}\BibitemShut
  {NoStop}%
\bibitem [{\citenamefont {Grosso}\ and\ \citenamefont
  {Parravicini}(2000)}]{grosso2000solid}%
  \BibitemOpen
  \bibfield  {author} {\bibinfo {author} {\bibfnamefont {G.}~\bibnamefont
  {Grosso}}\ and\ \bibinfo {author} {\bibfnamefont {G.}~\bibnamefont
  {Parravicini}},\ }\href {https://books.google.co.uk/books?id=L5RrQbbvWn8C}
  {\emph {\bibinfo {title} {Solid State Physics}}}\ (\bibinfo  {publisher}
  {Elsevier Science},\ \bibinfo {year} {2000})\BibitemShut {NoStop}%
\bibitem [{\citenamefont {Giuliani}\ and\ \citenamefont
  {Vignale}(2008)}]{giuliani2008quantum}%
  \BibitemOpen
  \bibfield  {author} {\bibinfo {author} {\bibfnamefont {G.}~\bibnamefont
  {Giuliani}}\ and\ \bibinfo {author} {\bibfnamefont {G.}~\bibnamefont
  {Vignale}},\ }\href {https://books.google.co.uk/books?id=FFydOyv\_r78C}
  {\emph {\bibinfo {title} {Quantum Theory of the Electron Liquid}}}\ (\bibinfo
   {publisher} {Cambridge University Press},\ \bibinfo {year}
  {2008})\BibitemShut {NoStop}%
\bibitem [{\citenamefont {Feibelman}(1968)}]{old_paper}%
  \BibitemOpen
  \bibfield  {author} {\bibinfo {author} {\bibfnamefont {P.~J.}\ \bibnamefont
  {Feibelman}},\ }\href@noop {} {\bibfield  {journal} {\bibinfo  {journal}
  {Phys. Rev.}\ }\textbf {\bibinfo {volume} {176}},\ \bibinfo {pages} {551}
  (\bibinfo {year} {1968})}\BibitemShut {NoStop}%
\bibitem [{\citenamefont {Ichikawa}(2011)}]{LP_in_RPA}%
  \BibitemOpen
  \bibfield  {author} {\bibinfo {author} {\bibfnamefont {M.}~\bibnamefont
  {Ichikawa}},\ }\href@noop {} {\bibfield  {journal} {\bibinfo  {journal} {J.
  Phys. Soc. Jpn.}\ }\textbf {\bibinfo {volume} {80}},\ \bibinfo {pages}
  {044606} (\bibinfo {year} {2011})}\BibitemShut {NoStop}%
\bibitem [{\citenamefont {Dasgupta}(1977)}]{dasgupta1977surface}%
  \BibitemOpen
  \bibfield  {author} {\bibinfo {author} {\bibfnamefont {B.~B.}\ \bibnamefont
  {Dasgupta}},\ }\href@noop {} {\bibfield  {journal} {\bibinfo  {journal} {Z.
  Phys. B}\ }\textbf {\bibinfo {volume} {27}},\ \bibinfo {pages} {75} (\bibinfo
  {year} {1977})}\BibitemShut {NoStop}%
\bibitem [{\citenamefont {Ekardt}(1984)}]{ekardt1984dynamical}%
  \BibitemOpen
  \bibfield  {author} {\bibinfo {author} {\bibfnamefont {W.}~\bibnamefont
  {Ekardt}},\ }\href@noop {} {\bibfield  {journal} {\bibinfo  {journal} {Phys.
  Rev. Lett.}\ }\textbf {\bibinfo {volume} {52}},\ \bibinfo {pages} {1925}
  (\bibinfo {year} {1984})}\BibitemShut {NoStop}%
\bibitem [{\citenamefont {Jin}(2015)}]{jin2015atomically}%
  \BibitemOpen
  \bibfield  {author} {\bibinfo {author} {\bibfnamefont {R.}~\bibnamefont
  {Jin}},\ }\href@noop {} {\bibfield  {journal} {\bibinfo  {journal}
  {Nanoscale}\ }\textbf {\bibinfo {volume} {7}},\ \bibinfo {pages} {1549}
  (\bibinfo {year} {2015})}\BibitemShut {NoStop}%
\bibitem [{\citenamefont {Weissker}\ \emph {et~al.}(2014)\citenamefont
  {Weissker}, \citenamefont {Escobar}, \citenamefont {Thanthirige},
  \citenamefont {Kwak}, \citenamefont {Lee}, \citenamefont {Ramakrishna},
  \citenamefont {Whetten},\ and\ \citenamefont
  {L{\'o}pez-Lozano}}]{weissker2014information}%
  \BibitemOpen
  \bibfield  {author} {\bibinfo {author} {\bibfnamefont {H.-C.}\ \bibnamefont
  {Weissker}}, \bibinfo {author} {\bibfnamefont {H.~B.}\ \bibnamefont
  {Escobar}}, \bibinfo {author} {\bibfnamefont {V.}~\bibnamefont
  {Thanthirige}}, \bibinfo {author} {\bibfnamefont {K.}~\bibnamefont {Kwak}},
  \bibinfo {author} {\bibfnamefont {D.}~\bibnamefont {Lee}}, \bibinfo {author}
  {\bibfnamefont {G.}~\bibnamefont {Ramakrishna}}, \bibinfo {author}
  {\bibfnamefont {R.}~\bibnamefont {Whetten}}, \ and\ \bibinfo {author}
  {\bibfnamefont {X.}~\bibnamefont {L{\'o}pez-Lozano}},\ }\href@noop {}
  {\bibfield  {journal} {\bibinfo  {journal} {Nat Commun}\ }\textbf {\bibinfo
  {volume} {5}},\ \bibinfo {pages} {3785} (\bibinfo {year} {2014})}\BibitemShut
  {NoStop}%
\bibitem [{\citenamefont {Philip}\ \emph {et~al.}(2012)\citenamefont {Philip},
  \citenamefont {Chantharasupawong}, \citenamefont {Qian}, \citenamefont
  {Jin},\ and\ \citenamefont {Thomas}}]{philip2012evolution}%
  \BibitemOpen
  \bibfield  {author} {\bibinfo {author} {\bibfnamefont {R.}~\bibnamefont
  {Philip}}, \bibinfo {author} {\bibfnamefont {P.}~\bibnamefont
  {Chantharasupawong}}, \bibinfo {author} {\bibfnamefont {H.}~\bibnamefont
  {Qian}}, \bibinfo {author} {\bibfnamefont {R.}~\bibnamefont {Jin}}, \ and\
  \bibinfo {author} {\bibfnamefont {J.}~\bibnamefont {Thomas}},\ }\href@noop {}
  {\bibfield  {journal} {\bibinfo  {journal} {Nano Lett.}\ }\textbf {\bibinfo
  {volume} {12}},\ \bibinfo {pages} {4661} (\bibinfo {year}
  {2012})}\BibitemShut {NoStop}%
\bibitem [{\citenamefont {Knoppe}\ \emph {et~al.}(2015)\citenamefont {Knoppe},
  \citenamefont {Vanbel}, \citenamefont {Van~Cleuvenbergen}, \citenamefont
  {Vanpraet}, \citenamefont {Bürgi},\ and\ \citenamefont
  {Verbiest}}]{knoppe2015nonlinear}%
  \BibitemOpen
  \bibfield  {author} {\bibinfo {author} {\bibfnamefont {S.}~\bibnamefont
  {Knoppe}}, \bibinfo {author} {\bibfnamefont {M.}~\bibnamefont {Vanbel}},
  \bibinfo {author} {\bibfnamefont {S.}~\bibnamefont {Van~Cleuvenbergen}},
  \bibinfo {author} {\bibfnamefont {L.}~\bibnamefont {Vanpraet}}, \bibinfo
  {author} {\bibfnamefont {T.}~\bibnamefont {Bürgi}}, \ and\ \bibinfo
  {author} {\bibfnamefont {T.}~\bibnamefont {Verbiest}},\ }\href@noop {}
  {\bibfield  {journal} {\bibinfo  {journal} {J. Phys. Chem. C}\ }\textbf
  {\bibinfo {volume} {119}},\ \bibinfo {pages} {6221} (\bibinfo {year}
  {2015})}\BibitemShut {NoStop}%
\bibitem [{\citenamefont {Knoppe}\ \emph {et~al.}(2016)\citenamefont {Knoppe},
  \citenamefont {Zhang}, \citenamefont {Wan}, \citenamefont {Wang},
  \citenamefont {Wang},\ and\ \citenamefont {Verbiest}}]{knoppe2016second}%
  \BibitemOpen
  \bibfield  {author} {\bibinfo {author} {\bibfnamefont {S.}~\bibnamefont
  {Knoppe}}, \bibinfo {author} {\bibfnamefont {Q.-F.}\ \bibnamefont {Zhang}},
  \bibinfo {author} {\bibfnamefont {X.-K.}\ \bibnamefont {Wan}}, \bibinfo
  {author} {\bibfnamefont {Q.-M.}\ \bibnamefont {Wang}}, \bibinfo {author}
  {\bibfnamefont {L.-S.}\ \bibnamefont {Wang}}, \ and\ \bibinfo {author}
  {\bibfnamefont {T.}~\bibnamefont {Verbiest}},\ }\href@noop {} {\bibfield
  {journal} {\bibinfo  {journal} {Ind. Eng. Chem. Res.}\ }\textbf {\bibinfo
  {volume} {55}},\ \bibinfo {pages} {10500} (\bibinfo {year}
  {2016})}\BibitemShut {NoStop}%
\bibitem [{\citenamefont {Yan}\ \emph {et~al.}(2007)\citenamefont {Yan},
  \citenamefont {Yuan},\ and\ \citenamefont {Gao}}]{yan2007end}%
  \BibitemOpen
  \bibfield  {author} {\bibinfo {author} {\bibfnamefont {J.}~\bibnamefont
  {Yan}}, \bibinfo {author} {\bibfnamefont {Z.}~\bibnamefont {Yuan}}, \ and\
  \bibinfo {author} {\bibfnamefont {S.}~\bibnamefont {Gao}},\ }\href@noop {}
  {\bibfield  {journal} {\bibinfo  {journal} {Phys. Rev. Lett.}\ }\textbf
  {\bibinfo {volume} {98}},\ \bibinfo {pages} {216602} (\bibinfo {year}
  {2007})}\BibitemShut {NoStop}%
\bibitem [{\citenamefont {Yasuike}\ \emph {et~al.}(2011)\citenamefont
  {Yasuike}, \citenamefont {Nobusada},\ and\ \citenamefont
  {Hayashi}}]{yasuike2011collectivity}%
  \BibitemOpen
  \bibfield  {author} {\bibinfo {author} {\bibfnamefont {T.}~\bibnamefont
  {Yasuike}}, \bibinfo {author} {\bibfnamefont {K.}~\bibnamefont {Nobusada}}, \
  and\ \bibinfo {author} {\bibfnamefont {M.}~\bibnamefont {Hayashi}},\
  }\href@noop {} {\bibfield  {journal} {\bibinfo  {journal} {Phys. Rev. A}\
  }\textbf {\bibinfo {volume} {83}},\ \bibinfo {pages} {013201} (\bibinfo
  {year} {2011})}\BibitemShut {NoStop}%
\bibitem [{\citenamefont {Bernadotte}\ \emph {et~al.}(2013)\citenamefont
  {Bernadotte}, \citenamefont {Evers},\ and\ \citenamefont
  {Jacob}}]{bernadotte2013plasmons}%
  \BibitemOpen
  \bibfield  {author} {\bibinfo {author} {\bibfnamefont {S.}~\bibnamefont
  {Bernadotte}}, \bibinfo {author} {\bibfnamefont {F.}~\bibnamefont {Evers}}, \
  and\ \bibinfo {author} {\bibfnamefont {C.~R.}\ \bibnamefont {Jacob}},\
  }\href@noop {} {\bibfield  {journal} {\bibinfo  {journal} {J. Phys. Chem. C}\
  }\textbf {\bibinfo {volume} {117}},\ \bibinfo {pages} {1863} (\bibinfo {year}
  {2013})}\BibitemShut {NoStop}%
\bibitem [{\citenamefont {Ma}\ \emph {et~al.}(2015)\citenamefont {Ma},
  \citenamefont {Wang},\ and\ \citenamefont {Wang}}]{ma2015interplay}%
  \BibitemOpen
  \bibfield  {author} {\bibinfo {author} {\bibfnamefont {J.}~\bibnamefont
  {Ma}}, \bibinfo {author} {\bibfnamefont {Z.}~\bibnamefont {Wang}}, \ and\
  \bibinfo {author} {\bibfnamefont {L.-W.}\ \bibnamefont {Wang}},\ }\href@noop
  {} {\bibfield  {journal} {\bibinfo  {journal} {Nat. Commun.}\ }\textbf
  {\bibinfo {volume} {6}},\ \bibinfo {pages} {10107} (\bibinfo {year}
  {2015})}\BibitemShut {NoStop}%
\bibitem [{\citenamefont {Govorov}\ \emph {et~al.}(2014)\citenamefont
  {Govorov}, \citenamefont {Zhang}, \citenamefont {Demir},\ and\ \citenamefont
  {Gun’ko}}]{Govorov_review}%
  \BibitemOpen
  \bibfield  {author} {\bibinfo {author} {\bibfnamefont {A.~O.}\ \bibnamefont
  {Govorov}}, \bibinfo {author} {\bibfnamefont {H.}~\bibnamefont {Zhang}},
  \bibinfo {author} {\bibfnamefont {H.~V.}\ \bibnamefont {Demir}}, \ and\
  \bibinfo {author} {\bibfnamefont {Y.~K.}\ \bibnamefont {Gun’ko}},\
  }\href@noop {} {\bibfield  {journal} {\bibinfo  {journal} {Nano Today}\
  }\textbf {\bibinfo {volume} {9}},\ \bibinfo {pages} {85} (\bibinfo {year}
  {2014})}\BibitemShut {NoStop}%
\bibitem [{\citenamefont {Zhang}\ and\ \citenamefont {Govorov}(2014)}]{Gov}%
  \BibitemOpen
  \bibfield  {author} {\bibinfo {author} {\bibfnamefont {H.}~\bibnamefont
  {Zhang}}\ and\ \bibinfo {author} {\bibfnamefont {A.~O.}\ \bibnamefont
  {Govorov}},\ }\href@noop {} {\bibfield  {journal} {\bibinfo  {journal} {J.
  Phys. Chem. C}\ }\textbf {\bibinfo {volume} {118}},\ \bibinfo {pages} {7606}
  (\bibinfo {year} {2014})}\BibitemShut {NoStop}%
\bibitem [{\citenamefont {Bursi}\ \emph {et~al.}(2016)\citenamefont {Bursi},
  \citenamefont {Calzolari}, \citenamefont {Corni},\ and\ \citenamefont
  {Molinari}}]{bursi2016quantifying}%
  \BibitemOpen
  \bibfield  {author} {\bibinfo {author} {\bibfnamefont {L.}~\bibnamefont
  {Bursi}}, \bibinfo {author} {\bibfnamefont {A.}~\bibnamefont {Calzolari}},
  \bibinfo {author} {\bibfnamefont {S.}~\bibnamefont {Corni}}, \ and\ \bibinfo
  {author} {\bibfnamefont {E.}~\bibnamefont {Molinari}},\ }\href@noop {}
  {\bibfield  {journal} {\bibinfo  {journal} {ACS Photonics}\ }\textbf
  {\bibinfo {volume} {3}},\ \bibinfo {pages} {520} (\bibinfo {year}
  {2016})}\BibitemShut {NoStop}%
\bibitem [{\citenamefont {Scholl}\ \emph {et~al.}(2012)\citenamefont {Scholl},
  \citenamefont {Koh},\ and\ \citenamefont {Dionne}}]{scholl2012quantum}%
  \BibitemOpen
  \bibfield  {author} {\bibinfo {author} {\bibfnamefont {J.~A.}\ \bibnamefont
  {Scholl}}, \bibinfo {author} {\bibfnamefont {A.~L.}\ \bibnamefont {Koh}}, \
  and\ \bibinfo {author} {\bibfnamefont {J.~A.}\ \bibnamefont {Dionne}},\
  }\href@noop {} {\bibfield  {journal} {\bibinfo  {journal} {Nature}\ }\textbf
  {\bibinfo {volume} {483}},\ \bibinfo {pages} {421} (\bibinfo {year}
  {2012})}\BibitemShut {NoStop}%
\bibitem [{\citenamefont {Zhang}\ \emph {et~al.}(2017)\citenamefont {Zhang},
  \citenamefont {Bursi}, \citenamefont {Cox}, \citenamefont {Cui},
  \citenamefont {Krauter}, \citenamefont {Alabastri}, \citenamefont
  {Manjavacas}, \citenamefont {Calzolari}, \citenamefont {Corni}, \citenamefont
  {Molinari}, \citenamefont {Carter}, \citenamefont {{Garc{\'{\i}}a De Abajo}},
  \citenamefont {Zhang},\ and\ \citenamefont {Nordlander}}]{Zhang2017}%
  \BibitemOpen
  \bibfield  {author} {\bibinfo {author} {\bibfnamefont {R.}~\bibnamefont
  {Zhang}}, \bibinfo {author} {\bibfnamefont {L.}~\bibnamefont {Bursi}},
  \bibinfo {author} {\bibfnamefont {J.~D.}\ \bibnamefont {Cox}}, \bibinfo
  {author} {\bibfnamefont {Y.}~\bibnamefont {Cui}}, \bibinfo {author}
  {\bibfnamefont {C.~M.}\ \bibnamefont {Krauter}}, \bibinfo {author}
  {\bibfnamefont {A.}~\bibnamefont {Alabastri}}, \bibinfo {author}
  {\bibfnamefont {A.}~\bibnamefont {Manjavacas}}, \bibinfo {author}
  {\bibfnamefont {A.}~\bibnamefont {Calzolari}}, \bibinfo {author}
  {\bibfnamefont {S.}~\bibnamefont {Corni}}, \bibinfo {author} {\bibfnamefont
  {E.}~\bibnamefont {Molinari}}, \bibinfo {author} {\bibfnamefont {E.~A.}\
  \bibnamefont {Carter}}, \bibinfo {author} {\bibfnamefont {F.~J.}\
  \bibnamefont {{Garc{\'{\i}}a De Abajo}}}, \bibinfo {author} {\bibfnamefont
  {H.}~\bibnamefont {Zhang}}, \ and\ \bibinfo {author} {\bibfnamefont
  {P.}~\bibnamefont {Nordlander}},\ }\href@noop {} {\bibfield  {journal}
  {\bibinfo  {journal} {ACS Nano}\ }\textbf {\bibinfo {volume} {11}},\ \bibinfo
  {pages} {7321} (\bibinfo {year} {2017})}\BibitemShut {NoStop}%
\bibitem [{\citenamefont {Sinha-Roy}\ \emph {et~al.}(2017)\citenamefont
  {Sinha-Roy}, \citenamefont {Garc{\'i}a-Gonz{\'a}lez}, \citenamefont
  {Weissker}, \citenamefont {Rabilloud},\ and\ \citenamefont
  {Fern{\'a}ndez-Dom{\'i}nguez}}]{sinha2017classical}%
  \BibitemOpen
  \bibfield  {author} {\bibinfo {author} {\bibfnamefont {R.}~\bibnamefont
  {Sinha-Roy}}, \bibinfo {author} {\bibfnamefont {P.}~\bibnamefont
  {Garc{\'i}a-Gonz{\'a}lez}}, \bibinfo {author} {\bibfnamefont {H.-C.}\
  \bibnamefont {Weissker}}, \bibinfo {author} {\bibfnamefont {F.}~\bibnamefont
  {Rabilloud}}, \ and\ \bibinfo {author} {\bibfnamefont {A.~I.}\ \bibnamefont
  {Fern{\'a}ndez-Dom{\'i}nguez}},\ }\href@noop {} {\bibfield  {journal}
  {\bibinfo  {journal} {ACS Photonics}\ }\textbf {\bibinfo {volume} {4}},\
  \bibinfo {pages} {1484} (\bibinfo {year} {2017})}\BibitemShut {NoStop}%
\bibitem [{\citenamefont {Cirac\`{\i}}\ and\ \citenamefont
  {Della~Sala}(2016)}]{ciraci2016quantum}%
  \BibitemOpen
  \bibfield  {author} {\bibinfo {author} {\bibfnamefont {C.}~\bibnamefont
  {Cirac\`{\i}}}\ and\ \bibinfo {author} {\bibfnamefont {F.}~\bibnamefont
  {Della~Sala}},\ }\href@noop {} {\bibfield  {journal} {\bibinfo  {journal}
  {Phys. Rev. B}\ }\textbf {\bibinfo {volume} {93}},\ \bibinfo {pages} {205405}
  (\bibinfo {year} {2016})}\BibitemShut {NoStop}%
\bibitem [{\citenamefont {Cirac\`{\i}}(2017)}]{ciraci2017current}%
  \BibitemOpen
  \bibfield  {author} {\bibinfo {author} {\bibfnamefont {C.}~\bibnamefont
  {Cirac\`{\i}}},\ }\href@noop {} {\bibfield  {journal} {\bibinfo  {journal}
  {Phys. Rev. B}\ }\textbf {\bibinfo {volume} {95}},\ \bibinfo {pages} {245434}
  (\bibinfo {year} {2017})}\BibitemShut {NoStop}%
\bibitem [{\citenamefont {Rossi}\ and\ \citenamefont {Kuhn}(2002)}]{Rossi2002}%
  \BibitemOpen
  \bibfield  {author} {\bibinfo {author} {\bibfnamefont {F.}~\bibnamefont
  {Rossi}}\ and\ \bibinfo {author} {\bibfnamefont {T.}~\bibnamefont {Kuhn}},\
  }\href@noop {} {\bibfield  {journal} {\bibinfo  {journal} {Rev. Mod. Phys.}\
  }\textbf {\bibinfo {volume} {74}},\ \bibinfo {pages} {895} (\bibinfo {year}
  {2002})}\BibitemShut {NoStop}%
\bibitem [{\citenamefont {Harris}(1984)}]{PhysRevA29}%
  \BibitemOpen
  \bibfield  {author} {\bibinfo {author} {\bibfnamefont {J.}~\bibnamefont
  {Harris}},\ }\href@noop {} {\bibfield  {journal} {\bibinfo  {journal} {Phys.
  Rev. A}\ }\textbf {\bibinfo {volume} {29}},\ \bibinfo {pages} {1648}
  (\bibinfo {year} {1984})}\BibitemShut {NoStop}%
\bibitem [{\citenamefont {Ernzerhof}(1996)}]{Ernzerhof1996}%
  \BibitemOpen
  \bibfield  {author} {\bibinfo {author} {\bibfnamefont {M.}~\bibnamefont
  {Ernzerhof}},\ }\href@noop {} {\bibfield  {journal} {\bibinfo  {journal}
  {Chem. Phys. Lett.}\ }\textbf {\bibinfo {volume} {263}},\ \bibinfo {pages}
  {499 } (\bibinfo {year} {1996})}\BibitemShut {NoStop}%
\bibitem [{\citenamefont {Rebolini}\ \emph {et~al.}(2014)\citenamefont
  {Rebolini}, \citenamefont {Toulouse}, \citenamefont {Teale}, \citenamefont
  {Helgaker},\ and\ \citenamefont {Savin}}]{ad_connection_2014}%
  \BibitemOpen
  \bibfield  {author} {\bibinfo {author} {\bibfnamefont {E.}~\bibnamefont
  {Rebolini}}, \bibinfo {author} {\bibfnamefont {J.}~\bibnamefont {Toulouse}},
  \bibinfo {author} {\bibfnamefont {A.~M.}\ \bibnamefont {Teale}}, \bibinfo
  {author} {\bibfnamefont {T.}~\bibnamefont {Helgaker}}, \ and\ \bibinfo
  {author} {\bibfnamefont {A.}~\bibnamefont {Savin}},\ }\href@noop {}
  {\bibfield  {journal} {\bibinfo  {journal} {J. Chem. Phys.}\ }\textbf
  {\bibinfo {volume} {141}},\ \bibinfo {eid} {044123} (\bibinfo {year}
  {2014})}\BibitemShut {NoStop}%
\bibitem [{\citenamefont {Rebolini}\ \emph {et~al.}(2015)\citenamefont
  {Rebolini}, \citenamefont {Toulouse}, \citenamefont {Teale}, \citenamefont
  {Helgaker},\ and\ \citenamefont {Savin}}]{PhysRevA91}%
  \BibitemOpen
  \bibfield  {author} {\bibinfo {author} {\bibfnamefont {E.}~\bibnamefont
  {Rebolini}}, \bibinfo {author} {\bibfnamefont {J.}~\bibnamefont {Toulouse}},
  \bibinfo {author} {\bibfnamefont {A.~M.}\ \bibnamefont {Teale}}, \bibinfo
  {author} {\bibfnamefont {T.}~\bibnamefont {Helgaker}}, \ and\ \bibinfo
  {author} {\bibfnamefont {A.}~\bibnamefont {Savin}},\ }\href@noop {}
  {\bibfield  {journal} {\bibinfo  {journal} {Phys. Rev. A}\ }\textbf {\bibinfo
  {volume} {91}},\ \bibinfo {pages} {032519} (\bibinfo {year}
  {2015})}\BibitemShut {NoStop}%
\bibitem [{\citenamefont {Novotny}\ and\ \citenamefont
  {Hecht}(2012)}]{nano-optics}%
  \BibitemOpen
  \bibfield  {author} {\bibinfo {author} {\bibfnamefont {L.}~\bibnamefont
  {Novotny}}\ and\ \bibinfo {author} {\bibfnamefont {B.}~\bibnamefont
  {Hecht}},\ }\href@noop {} {\emph {\bibinfo {title} {Principles of
  Nano-Optics}}},\ \bibinfo {edition} {2nd}\ ed.\ (\bibinfo  {publisher}
  {Cambridge University Press},\ \bibinfo {year} {2012})\BibitemShut {NoStop}%
\bibitem [{\citenamefont {Slepyan}\ \emph {et~al.}(2004)\citenamefont
  {Slepyan}, \citenamefont {Magyarov}, \citenamefont {Maksimenko},
  \citenamefont {Hoffmann},\ and\ \citenamefont {Bimberg}}]{Slepyan2004}%
  \BibitemOpen
  \bibfield  {author} {\bibinfo {author} {\bibfnamefont {G.~Y.}\ \bibnamefont
  {Slepyan}}, \bibinfo {author} {\bibfnamefont {A.}~\bibnamefont {Magyarov}},
  \bibinfo {author} {\bibfnamefont {S.~A.}\ \bibnamefont {Maksimenko}},
  \bibinfo {author} {\bibfnamefont {A.}~\bibnamefont {Hoffmann}}, \ and\
  \bibinfo {author} {\bibfnamefont {D.}~\bibnamefont {Bimberg}},\ }\href@noop
  {} {\bibfield  {journal} {\bibinfo  {journal} {Phys. Rev. B}\ }\textbf
  {\bibinfo {volume} {70}},\ \bibinfo {pages} {045320} (\bibinfo {year}
  {2004})}\BibitemShut {NoStop}%
\bibitem [{\citenamefont {Ruggenthaler}\ and\ \citenamefont
  {Bauer}(2009)}]{ruggenthaler2009rabi}%
  \BibitemOpen
  \bibfield  {author} {\bibinfo {author} {\bibfnamefont {M.}~\bibnamefont
  {Ruggenthaler}}\ and\ \bibinfo {author} {\bibfnamefont {D.}~\bibnamefont
  {Bauer}},\ }\href@noop {} {\bibfield  {journal} {\bibinfo  {journal} {Phys.
  Rev. Lett.}\ }\textbf {\bibinfo {volume} {102}},\ \bibinfo {pages} {233001}
  (\bibinfo {year} {2009})}\BibitemShut {NoStop}%
\bibitem [{\citenamefont {Fuks}\ \emph {et~al.}(2011)\citenamefont {Fuks},
  \citenamefont {Helbig}, \citenamefont {Tokatly},\ and\ \citenamefont
  {Rubio}}]{fuks2011nonlinear}%
  \BibitemOpen
  \bibfield  {author} {\bibinfo {author} {\bibfnamefont {J.~I.}\ \bibnamefont
  {Fuks}}, \bibinfo {author} {\bibfnamefont {N.}~\bibnamefont {Helbig}},
  \bibinfo {author} {\bibfnamefont {I.~V.}\ \bibnamefont {Tokatly}}, \ and\
  \bibinfo {author} {\bibfnamefont {A.}~\bibnamefont {Rubio}},\ }\href@noop {}
  {\bibfield  {journal} {\bibinfo  {journal} {Phys. Rev. B}\ }\textbf {\bibinfo
  {volume} {84}},\ \bibinfo {pages} {075107} (\bibinfo {year}
  {2011})}\BibitemShut {NoStop}%
\bibitem [{\citenamefont {Provorse}\ and\ \citenamefont
  {Isborn}(2016)}]{provorse2016electron}%
  \BibitemOpen
  \bibfield  {author} {\bibinfo {author} {\bibfnamefont {M.~R.}\ \bibnamefont
  {Provorse}}\ and\ \bibinfo {author} {\bibfnamefont {C.~M.}\ \bibnamefont
  {Isborn}},\ }\href@noop {} {\bibfield  {journal} {\bibinfo  {journal} {Int.
  J. Quantum Chem.}\ }\textbf {\bibinfo {volume} {116}},\ \bibinfo {pages}
  {739} (\bibinfo {year} {2016})}\BibitemShut {NoStop}%
\bibitem [{\citenamefont {Haug}\ and\ \citenamefont
  {Koch}(2009)}]{haug2009quantum}%
  \BibitemOpen
  \bibfield  {author} {\bibinfo {author} {\bibfnamefont {H.}~\bibnamefont
  {Haug}}\ and\ \bibinfo {author} {\bibfnamefont {S.~W.}\ \bibnamefont
  {Koch}},\ }\href@noop {} {\emph {\bibinfo {title} {Quantum Theory of the
  Optical and Electronic Properties of Semiconductors: Fifth Edition}}}\
  (\bibinfo  {publisher} {World Scientific Publishing Company},\ \bibinfo
  {year} {2009})\BibitemShut {NoStop}%
\end{thebibliography}

%

\end{document}